\documentclass[conference]{IEEEtran}
\IEEEoverridecommandlockouts
\usepackage{cite}
\usepackage{amsmath,amssymb,amsfonts}
\usepackage{subfig}
\usepackage{algorithmic}
\usepackage{algorithm}
\usepackage{graphicx}
\usepackage{textcomp}
\usepackage{xcolor}
\usepackage{tikz}
\usepackage{caption}
\usetikzlibrary{shapes.geometric, arrows.meta, positioning, calc, backgrounds, fit}

\def\BibTeX{{\rm B\kern-.05em{\sc i\kern-.025em b}\kern-.08em
    T\kern-.1667em\lower.7ex\hbox{E}\kern-.125emX}}

\begin{document}

\title{Out-of-Distribution Detection in Wireless
Multimodal Foundation Models for 6G ISAC \vspace{-1ex}
\thanks{This work has been supported by MITACS and Ericsson Canada.}
}

\author{
\normalsize
Mohammad Farzanullah\IEEEauthorrefmark{1}, Akram Bin Sediq\IEEEauthorrefmark{2}, Ali Afana\IEEEauthorrefmark{2} and Melike Erol-Kantarci\IEEEauthorrefmark{1}\\
\vspace{0ex}
\IEEEauthorblockA{\IEEEauthorrefmark{1}
    School of Electrical Engineering and Computer Science, University of Ottawa, Ottawa, ON, Canada} 
\IEEEauthorblockA{\IEEEauthorrefmark{2}
    Ericsson Inc., Ottawa, ON, Canada}
Emails: \{mfarz086, melike.erolkantarci\}@uottawa.ca, \{akram.bin.sediq, ali.afana\}@ericsson.com
\vspace{-2ex}
}

\maketitle

\begin{abstract}
The integration of Foundation Models (FMs), such as the Wireless Multimodal Foundation Model (WMFM), into 6G networks provides a unified framework for Integrated Sensing and Communication (ISAC), leveraging generalized representations to simultaneously optimize data transmission and environmental perception.
However, the deployment of such data-driven models in safety-critical infrastructure is hindered by the Out-of-Distribution (OOD) problem, which poses a fundamental threat to system trustworthiness. Standard FMs operate under a closed-world assumption, rendering them vulnerable to silent failures when deployed in unseen radio environments. To address this reliability gap and ensure trustworthy network operation, we propose \textbf{WMFM-OOD}, a robust metric-based OOD detection framework. Unlike traditional methods that rely on raw compatibility scores, WMFM-OOD constructs geometric Base Station (BS) Prototypes within the joint latent space to capture the manifold structure of valid radio environments. By employing a temperature-scaled probabilistic scoring mechanism, our approach effectively distinguishes between In-Distribution (ID) and covariate-shifted anomalies. We validate the framework on the DeepVerse6G dataset. Experimental results demonstrate that WMFM-OOD significantly outperforms uncalibrated baselines, achieving an Area Under the Receiver Operating Characteristic Curve (AUROC) of 0.8824 and reducing the False Positive Rate (FPR)  at 95 \% True Positive Rate (TPR), commonly referred to as FPR95, by approximately 17\% in the optimal temperature regime, thereby providing an initial layer of detection sensitivity to mitigate catastrophic model failures without completely disrupting network availability.
\end{abstract}


\vspace{-2ex}
\section{Introduction}
The evolution of wireless networks toward the sixth generation (6G) envisions an artificial intelligence (AI)-native air interface, where main physical layer functionalities, such as channel estimation, beamforming, and positioning, are increasingly driven by Deep Neural Networks (DNNs) rather than rigid analytical models. 
This paradigm shift is particularly critical for Integrated Sensing and Communication (ISAC) systems. Unlike traditional networks, ISAC leverages a unified hardware and spectrum framework to perform dual functions: high-speed data transmission and high-resolution environmental sensing (e.g., radar-like target detection and localization). In such complex scenarios, the integration of multimodal data is indispensable. For instance, incorporating auxiliary visual context from cameras can compensate for the inherent sparsity and noise of wireless channels. This multimodal synergy enables robust environmental understanding and precise beam alignment where Radio Frequency (RF)-only methods typically fail \cite{mm}.

Recently, the emergence of Foundation Models (FMs) in the wireless domain has accelerated this transition \cite{FM_survey}. 
These models serve as versatile, pre-trained architectures capable of capturing the complex, non-linear patterns of the electromagnetic environment across various frequency bands and deployment scenarios. By learning generalized representations of the wireless channel, FMs can be efficiently fine-tuned for a wide array of downstream tasks, such as CSI feedback compression, predictive beamforming, and interference management, often requiring significantly less site-specific data than traditional deep learning approaches.
For instance, the Wireless Multimodal Foundation Model (WMFM) \cite{WMFM} leverages contrastive learning to align RF signals with visual modalities, creating a shared latent representation that supports diverse downstream tasks with high data efficiency.

However, reliance on aligned cross-modal representations creates distinct vulnerabilities when facing the Out-of-Distribution (OOD) problem~\cite{borlino2024foundation}. 
Standard DNNs operate under the closed-world assumption, where training and testing data are drawn from identical distributions. In real-world 6G deployments, however, this assumption is routinely violated \cite{OOD_survey}. 
Practical environments are continuously altered by dynamic factors such as shifting channel statistics, user mobility, and evolving network topologies. 
Consequently, pre-trained FMs may encounter novel scattering profiles where learned cross-modal correlations degrade.
Despite this alignment collapse, standard FMs often fail to quantify the resulting model uncertainty, yielding highly confident but degenerate embeddings in unseen domains~\cite{amodei2016concrete}. 
Thus, integrating robust OOD detection is imperative to identify alignment breaches, ensuring the model abstains from inference outside its valid regime.

While extensive in computer vision, OOD detection in wireless is nascent; existing works primarily rely on unimodal RF features and generative reconstruction errors \cite{OODinWireless,OOD_in_wireless_second, onyekwelu2025out}.
Furthermore, general OOD literature often focuses on \textit{semantic shift}—the appearance of entirely new classes (e.g., a dog appearing in a cat classifier). In contrast, wireless network generalization is predominantly challenged by \textit{covariate shift} \cite{OODinWireless}. 
Here, environmental variations drastically alter the statistical distribution of input covariates (e.g., channel impulse responses) rather than the class labels.
Consequently, the model cannot rely on expanding its label vocabulary; it must instead capture the underlying structural variations of the environment itself.
Addressing covariate shift is essential for wireless FMs, which must maintain robust feature representations across diverse, unseen geographic deployments.

In this paper, we propose \textbf{WMFM-OOD}, a metric-based OOD detection framework designed as a preliminary step toward addressing the reliability challenges of wireless multimodal foundation models. Building upon the WMFM architecture~\cite{WMFM}, our work bridges the gap between foundation model generalization and safety-critical deployment through three key contributions:
\begin{itemize}
    \item We adapt the Maximum Concept Matching (MCM) principle, originally designed for vision-language models~\cite{MCM}, to the wireless domain. Unlike standard MCM which relies on textual prompts, we formulate a geometric approach that utilizes aligned camera and Channel State Information (CSI) representations to detect distribution shifts without requiring retraining.
    
    \item We introduce the construction of ``Base Station (BS) Prototypes" within the joint latent space. By aggregating embeddings from known environments, we capture the unique manifold signatures of valid radio scenarios, enabling the model to distinguish between legitimate variations and fundamental environmental anomalies.
    
    \item We incorporate a temperature-scaled Softmax mechanism to refine detection sensitivity. This calibration sharpens the distinction between In-Distribution (ID) and OOD samples. By doing so, it avoids the risk of the model misclassifying OOD data, a frequent issue when using raw geometric distances without proper scaling.
\end{itemize}

We validate this approach using the DeepVerse6G dataset.
Our results demonstrate that WMFM-OOD significantly outperforms uncalibrated baselines, achieving an AUROC of 0.8824, and reducing the False Positive Rate at 95\% True Positive Rate (FPR95) by approximately 17\%, as compared to the benchmark, in the optimal temperature regime.

\section{System Model}

The investigated ISAC framework consists of $K$ BSs (denoted by set $\mathcal{K}$) featuring a uniform array of $M$ antennas and integrated RGB cameras for visual sensing. The network serves $U$ single-antenna User Equipments (UEs). 

In the uplink channel estimation stage, pilot symbols are transmitted over an OFDM waveform with $S$ subcarriers. The signal vector $\mathbf{r}_s \in \mathbb{C}^{M}$ received at the BS on the $s$-th subcarrier is modeled as:
\begin{equation}
    \mathbf{r}_s = \mathbf{h}_s p_s + \mathbf{n}_s, \quad s = 1, \dots, S,
\end{equation}
where $p_s$ is the transmitted pilot symbol, $\mathbf{n}_s \sim \mathcal{CN}(\mathbf{0}, \sigma^2\mathbf{I})$ denotes the additive Gaussian noise, and $\mathbf{h}_s \in \mathbb{C}^M$ represents the frequency-domain channel vector.

The aggregate spatial-frequency channel estimate is represented by the matrix $\mathbf{H} \in \mathbb{C}^{M \times S}$, constructed by concatenating the individual subcarrier vectors:
\begin{equation}
    \mathbf{H} = [\mathbf{h}_1, \mathbf{h}_2, \dots, \mathbf{h}_S].
\end{equation}
This matrix serves as the CSI input for the proposed model. 

To facilitate multimodal learning, the BS camera captures an image, $\mathbf{V}$, under the assumption that it is perfectly synchronized with the radio frame arrival. Consequently, the training dataset consists of tuples $\mathcal{T} = \{(\mathbf{H}, \mathbf{V}, y)\}$, where $y$ denotes the ground truth label for the target application (e.g., user positioning or LOS detection). This structure allows for the extraction of shared features between the radio frequency and visual domains.

\subsection{Problem Formulation}

The core objective of the multimodal framework in \cite{WMFM} is to learn parameterized modality-specific encoders, denoted as $f_{\text{rf}}(\cdot; \theta_{\text{rf}})$ and $f_{\text{vis}}(\cdot; \theta_{\text{vis}})$, which map the input data into a shared latent representation space. For the $i$-th sample, the latent feature vectors are obtained via:
\begin{equation}
    \mathbf{l}_{\text{rf}, i} = f_{\text{rf}}(\mathbf{H}_i), \quad \mathbf{l}_{\text{vis}, i} = f_{\text{vis}}(\mathbf{V}_i).
\end{equation}

The baseline training objective seeks to minimize a contrastive representation learning loss, $\mathcal{L}_{\text{rep}}$, enabling the alignment of these heterogeneous modalities into a shared latent space by maximizing the similarity of positive pairs \cite{WMFM}:
\begin{equation}
    \min_{\theta_{\text{rf}}, \theta_{\text{vis}}} \mathcal{L}_{\text{rep}}(\mathbf{l}_{\text{rf}}, \mathbf{l}_{\text{vis}}).
\end{equation}

Crucially, this work departs from the standard closed-set assumption and addresses a realistic \textit{open-set} scenario. We assume the presence of OOD data during deployment, implying a fundamental distribution shift between the training and testing environments, expressed as $\mathbb{P}_{\text{train}}(\mathbf{x}, y) \neq \mathbb{P}_{\text{test}}(\mathbf{x}, y)$.

Formally, let $\mathcal{X}_{\text{ID}}$ represent the set of ID samples and $\mathcal{X}_{\text{OOD}}$ represent the set of OOD samples. The training dataset, $\mathcal{D}_{\text{train}}$, is strictly composed of ID data:
\begin{equation}
    \mathcal{D}_{\text{train}} = \{(\mathbf{x}^{\text{tr}}_i, y^{\text{tr}}_i) \mid \mathbf{x}^{\text{tr}}_i \in \mathcal{X}_{\text{ID}}\}.
\end{equation}

Conversely, the test dataset, $\mathcal{D}_{\text{test}}$, encompasses a mixture of known and unknown distributions:
\begin{equation}
    \mathcal{D}_{\text{test}} = \{(\mathbf{x}^{\text{te}}_i, y^{\text{te}}_i) \mid \mathbf{x}^{\text{te}}_i \in \mathcal{X}_{\text{ID}} \cup \mathcal{X}_{\text{OOD}}\}.
\end{equation}

The main objective is to design a scoring function $S(\mathbf{x}; \theta): \mathcal{X} \to \mathbb{R}$ that assigns a scalar confidence score to any input sample $\mathbf{x}$. The OOD detection task is then defined as a binary classification problem with the decision rule:
\begin{equation}
    G(\mathbf{x}) = 
    \begin{cases} 
    \text{ID}, & \text{if } S(\mathbf{x}) \geq \gamma \\
    \text{OOD}, & \text{if } S(\mathbf{x}) < \gamma
    \end{cases},
\end{equation}
where $\gamma$ is a decision threshold. An optimal detector maximizes the separation between the scores of ID samples and OOD samples, ensuring that $\mathbb{E}_{\mathbf{x} \in \mathcal{X}_{\text{ID}}}[S(\mathbf{x})] \gg \mathbb{E}_{\mathbf{x} \in \mathcal{X}_{\text{OOD}}}[S(\mathbf{x})]$.

\section{Proposed Methodology: Metric-Based OOD Detection via Multimodal Concept Matching}

\begin{figure*}
    \centering
    \includegraphics[width=0.8\linewidth, trim=20pt 10pt 20pt 10pt, clip]{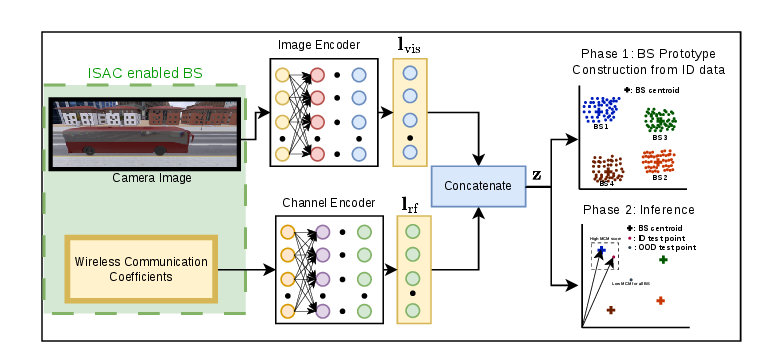}
    \caption{Architectural overview of the WMFM-OOD framework. Modality-specific features ($\mathbf{l}_{\text{vis}}$ and $\mathbf{l}_{\text{rf}}$) are extracted from synchronized camera frames and CSI matrices, fused via concatenation, and projected onto a unit hypersphere ($\textbf{z}$). Phase 1 (Offline) aggregates in-distribution data to construct stable geometric Base Station Prototypes ($\mu_k$). Phase 2 (Online Inference) evaluates incoming test points against these prototypes using a temperature-scaled Softmax to isolate environmental anomalies through directional uncertainty.}
    \label{fig:framework}
\end{figure*}

To tackle the open-set challenge outlined in Section II, we propose a robust metric-based OOD detection framework tailored for wireless environments. Our strategy adapts the MCM principle \cite{MCM}, originally designed for vision-language models, to the domain of multimodal wireless sensing. Unlike standard MCM, which relies on pre-defined textual prompts to form concepts, our approach constructs ``BS Prototypes" directly from the geometric structure of the learned latent space. This effectively captures the unique signature of each radio environment. The framework comprises three key stages: Feature Fusion, BS Prototype Construction, and Temperature-Scaled Scoring.

\subsection{Multimodal Feature Fusion}
Let $f_{\text{rf}}(\cdot)$ and $f_{\text{vis}}(\cdot)$ represent the pre-trained encoders for the RF and visual modalities, respectively. For a given sample, we extract modality-specific embeddings $\mathbf{l}_{\text{rf}} \in \mathbb{R}^{d_1}$ and $\mathbf{l}_{\text{vis}} \in \mathbb{R}^{d_2}$. To harness the complementary information from both domains, we synthesize a raw composite embedding via concatenation: $\mathbf{z}_{\text{raw}} = [\mathbf{l}_{\text{rf}}; \mathbf{l}_{\text{vis}}] \in \mathbb{R}^{d_1 + d_2}$.

To prevent magnitude disparities from skewing the similarity metrics, we enforce a spherical embedding space by projecting the composite vector onto the unit hypersphere:

\begin{equation}
    \mathbf{z} = \frac{\mathbf{z}_{\text{raw}}}{\|\mathbf{z}_{\text{raw}}\|_2}.
\end{equation}

This normalized vector $\mathbf{z}$ constitutes the fundamental input for our detection mechanism.

\subsection{BS Prototype Construction}
A distinct feature of our method is the offline construction of prototypes that explicitly represent valid BS environments. Utilizing the labeled In-Distribution ID training set $\mathcal{D}_{\text{train}}$, we generate a representative anchor for each known BS class $k \in \mathcal{K} = \{1, \dots, K\}$. Each BS corresponds to a distinct physical location.

For every BS class $k$, we aggregate the set of associated training embeddings $\mathcal{J}_k = \{i \mid y_i = k\}$. The class centroid is computed by averaging these normalized embeddings:

\begin{equation}
    \boldsymbol{\mu}_k^{\text{raw}} = \frac{1}{|\mathcal{J}_k|} \sum_{i \in \mathcal{J}_k} \mathbf{z}_i.
\end{equation}

To maintain consistency with the spherical input space, the final BS prototype $\boldsymbol{\mu}_k$ is obtained by re-normalizing the centroid:

\begin{equation}
    \boldsymbol{\mu}_k = \frac{\boldsymbol{\mu}_k^{\text{raw}}}{\|\boldsymbol{\mu}_k^{\text{raw}}\|_2}.
\end{equation}

These prototypes $\{\boldsymbol{\mu}_k\}_{k=1}^K$ serve as stable reference points, encapsulating the ideal multimodal features of each known BS environment
This prototype construction fundamentally departs from vision-language MCM \cite{MCM}. While MCM relies on discrete textual prompts, wireless channels lack linguistic equivalents for multipath scattering profiles. Consequently, we replace textual concepts with continuous, multi-sensor hyperspherical manifolds derived from synchronized radio-visual streams, transforming a semantic search into a localized geometric anchor system tailored for wireless environments.

\subsection{Temperature-Scaled MCM Scoring}
For an incoming test sample with normalized embedding $\mathbf{z}_{\text{test}}$, we evaluate its conformity to the known BS manifolds. We first compute the cosine similarity between the test sample and each BS prototype:

\begin{equation}
    s_k(\mathbf{z}_{\text{test}}) = \mathbf{z}_{\text{test}}^\top \boldsymbol{\mu}_k, \quad k \in \mathcal{K}.
\end{equation}

To amplify the separation between ID and OOD samples, we employ a temperature-scaled Softmax function with a parameter $\tau > 0$. The probability of the sample belonging to BS environment $k$ is given by:

\begin{equation}
    p_k(\mathbf{z}_{\text{test}}; \tau) = \frac{\exp(s_k(\mathbf{z}_{\text{test}}) / \tau)}{\sum_{j=1}^K \exp(s_j(\mathbf{z}_{\text{test}}) / \tau)}.
\end{equation}

The final OOD detection score is defined as the maximum probability across all known BS classes:

\begin{equation}
    S_{\text{MCM}}(\mathbf{z}_{\text{test}}) = \max_{k \in \mathcal{K}} p_k(\mathbf{z}_{\text{test}}; \tau).
\end{equation}

It is important to note that the inference process is label-agnostic. The scoring function $S_{MCM}$ evaluates the test sample against all constructed prototypes $\{ \mu_k \}_{k=1}^K$ without requiring knowledge of the ground truth BS ID. This ensures applicability in open-set scenarios where the true environmental class of an incoming signal may be unknown or undefined.
ID samples will typically align closely with their corresponding BS prototype, yielding $S_{\text{MCM}} \approx 1$. In contrast, OOD samples—originating from unknown environments—will exhibit low, uniform similarity across all prototypes, resulting in a significantly attenuated score.

The complete training and inference procedure is formalized in Algorithm 1, and depicted in Fig. \ref{fig:framework}.

\begin{algorithm}
\caption{Proposed WMFM-OOD framework}
\begin{algorithmic}[1]
\REQUIRE ID Training set $\mathcal{D}_{\text{train}}$, Pre-trained encoders $f_{\text{rf}}, f_{\text{vis}}$, Temperature $\tau$, Threshold $\gamma$.
\STATE \textbf{// Phase 1: BS Prototype Construction (Offline)}
\FOR{each BS class $k \in \{1, \dots, K\}$ in ID data}
    \STATE Collect embeddings $\mathcal{Z}_k = \{\mathbf{z}_i \mid y_i = k\}$ from $\mathcal{D}_{\text{train}}$
    \STATE Compute centroid: $\boldsymbol{\mu}_k^{\text{raw}} \leftarrow \text{Mean}(\mathcal{Z}_k)$
    \STATE Normalize prototype: $\boldsymbol{\mu}_k \leftarrow \boldsymbol{\mu}_k^{\text{raw}} / \|\boldsymbol{\mu}_k^{\text{raw}}\|_2$
\ENDFOR
\STATE \textbf{// Phase 2: Inference (Online)}
\FOR{each test sample $(\mathbf{H}, \mathbf{V})$}
    \STATE Extract features: $\mathbf{l}_{\text{rf}} \leftarrow f_{\text{rf}}(\mathbf{H}), \mathbf{l}_{\text{vis}} \leftarrow f_{\text{vis}}(\mathbf{V})$
    \STATE Fuse: $\mathbf{z}_{\text{raw}} \leftarrow [\mathbf{l}_{\text{rf}}; \mathbf{l}_{\text{vis}}]$
    \STATE Normalize: $\mathbf{z} \leftarrow \mathbf{z}_{\text{raw}} / \|\mathbf{z}_{\text{raw}}\|_2$
    \STATE Calculate similarities: $s_k \leftarrow \mathbf{z}^\top \boldsymbol{\mu}_k$ for all $k$
    \STATE Compute probabilities: $p_k \leftarrow \text{Softmax}(s_k / \tau)$
    \STATE Derive score: $S_{\text{MCM}} \leftarrow \max_k p_k$
    \IF{$S_{\text{MCM}} < \gamma$}
        \STATE Prediction $\leftarrow$ \textbf{OOD}
    \ELSE
        \STATE Prediction $\leftarrow$ \textbf{ID}
    \ENDIF
\ENDFOR
\end{algorithmic}
\end{algorithm}

The computational overhead of WMFM-OOD is negligible during the inference phase. Since the prototypes are pre-computed offline, the online scoring mechanism requires only vector dot products, scaling linearly as $\mathcal{O}(K \cdot d_{total})$ where $K$ is the number of BS classes and $d_{total}$ is the joint embedding dimension. This ensures the OOD detector adds virtually no latency to the backbone encoders, preserving the real-time capability required for 6G control loops.

\section{Simulation Settings}

\subsection{Simulation Setup}

To validate our proposed approach, we utilize the DeepVerse6G dataset~\cite{DeepVerse}. 
DeepVerse6G is a large-scale, simulated multimodal dataset that employs ray-tracing  methodologies to replicate physically accurate wireless propagation environments  alongside synchronized sensor data. Specifically, we select the O1 scenario as the ID dataset and the Carla-Town1 scenario as the OOD dataset. 
The O1 scenario simulates a highway environment featuring four BSs  positioned along the highway. Each BS is equipped with an antenna array for communication and three cameras providing panoramic visual coverage. 
In contrast, the Carla-Town1 scenario represents an urban, and residential intersection environment, featuring five BSs distributed across various intersections. Each BS is equipped with cameras providing full panoramic coverage.
Crucially, despite the macro-level distinctiveness of these geographic layouts, the underlying local radio-visual sub-manifolds exhibit severe feature aliasing, creating a highly complex covariate shift benchmark where simple nearest-neighbor or density boundary metrics suffer from severe distribution overlap.

We select the structurally distinct highway (O1) and urban (Carla-Town1) scenarios to maximize domain discrepancy, creating a challenging OOD benchmark. This setup is crucial for assessing FM reliability in unseen geographic deployments.

\subsection{Implementation details}
We employ the pretrained WMFM~\cite{WMFM} 
as the backbone for our validation. The WMFM utilizes a contrastive objective to project camera inputs and CSI into a unified latent space, thereby ensuring alignment between the two modalities. 
Crucially, the reliability of the WMFM relies on the assumption that this learned cross-modal alignment holds during inference. In OOD scenarios, such as unseen urban topologies, the relationship between visual and wireless features often degrades. 
Consequently, an effective OOD detection mechanism is essential to flag these alignment deviations, ensuring that downstream tasks do not rely on misaligned or degenerate embeddings.

To evaluate our OOD detection framework, we adopt standard metrics established in the literature~\cite{benchmark}: AUROC and FPR95. 
The \textbf{AUROC} offers a threshold-independent measure of separability, representing the probability that a random OOD sample receives a higher anomaly score than a random ID sample (where 50\% indicates random guessing). 
Complementing this, \textbf{FPR95} measures the percentage of OOD samples incorrectly classified as in-distribution when the detection threshold is tuned to maintain a 95\% True Positive Rate for ID data. Minimizing FPR95 is critical for wireless networks to ensure robust security against OOD signals while maintaining high service availability for legitimate links.
Finally, to assess the robustness of our framework across different scaling regimes, we perform a sensitivity analysis by varying the temperature parameter $\tau \in \{0.1, 0.2, 0.5, 1, 5, 10, 50, 100\}$.

To rigorously evaluate the efficacy of our proposed temperature-scaled scoring, we compare it against four baselines. 

\subsubsection{Baseline 1: Direct Modality Alignment ($S_{\text{DMA}}$)}
This baseline assesses the consistency between the two modalities directly, independent of the base station prototypes. We compute the dot product between the normalized camera embedding $\mathbf{l}_{\text{vis}}$ and the CSI embedding $\mathbf{l}_{\text{rf}}$ for the same sample:
\begin{equation}
    S_{\text{DMA}} = \mathbf{l}_{\text{vis}}^\top \mathbf{l}_{\text{rf}}.
\end{equation}
Since the WMFM is trained via a contrastive objective to maximize the similarity of matched pairs, a high dot product indicates the strong cross-modal coherence typical of ID data. Conversely, OOD samples, characterized by unseen environmental features, are expected to exhibit weaker alignment, resulting in lower scores. 

\subsubsection{Baseline 2: Maximum Prototype Similarity ($S_{\text{MPS}}$)}
This baseline evaluates the conformity of the test sample to the known BS manifolds using raw geometric distance, identical to the first step of our proposed method but without temperature scaling or normalization. For a test sample $\mathbf{z}_{\text{test}}$, we define the score as the maximum cosine similarity to any learned BS prototype:
\begin{equation}
    S_{\text{MPS}} = \max_{k \in \mathcal{K}} (\mathbf{z}_{\text{test}}^\top \boldsymbol{\mu}_k).
\end{equation}
This metric serves as an ablation study for the Softmax operation, allowing us to isolate the specific contribution of probabilistic calibration (and the temperature parameter $\tau$) toward OOD detection performance.

\subsubsection{Baseline 3: Energy-Based OOD ($S_{\text{Energy}}$):} To evaluate our framework against a non-probabilistic baseline, we adapt the energy-based OOD detection framework \cite{liu2020energy} to our logit-free architecture. Instead of forcing a bounded softmax probability distribution, the negative free energy is computed directly over the raw geometric prototype projections:
\begin{equation}
S_{\text{Energy}} = T_{\text{e}} \cdot \log \sum_{k=1}^{K} \exp\left(\frac{\mathbf{z}_{\text{test}}^\top \mu_k}{T_{\text{e}}}\right)
\end{equation}
where $T_{\text{e}}$ is the energy temperature hyperparameter. Lower energy scores imply a higher density conformity to the ID manifold, whereas anomalies exhibit structurally lower metrics. We conduct empirical evaluations by sweeping the energy temperature hyperparameter across the range $T_{\text{e}} \in [0.001, 100]$; for brevity, we report only the optimal performance achieved at $T_{\text{e}} = 0.01$.

\subsubsection{Baseline 4: Multimodal Modality Discrepancy ($S_{\text{MMD}}$):} This baseline evaluates the internal alignment consistency between the constituent modalities independent of environmental prototypes. Leveraging the frozen representation space, we compute the negative Euclidean distance between the normalized visual embedding $\mathbf{l}_{\text{vis}}$ and the radio frequency embedding $\mathbf{l}_{\text{rf}}$ for a given sample:
\begin{equation}
S_{\text{MMD}} = - \| f_{\text{vis}}(V) - f_{\text{rf}}(H) \|_2
\end{equation}
A severe covariate shift degrades the learned cross-modal alignment, manifesting as a wider distance (lower $S_{\text{MMD}}$ score), whereas in-distribution inputs are characterized by highly coherent, tightly aligned feature pairs.

\section{Results and Discussion}

In this section, we evaluate WMFM-OOD on the DeepVerse6G dataset, analyzing the effects of temperature scaling and modality fusion on detection robustness.
\begin{figure}
    \centering
    \subfloat[AUROC]{%
        \includegraphics[width=1\linewidth, clip]{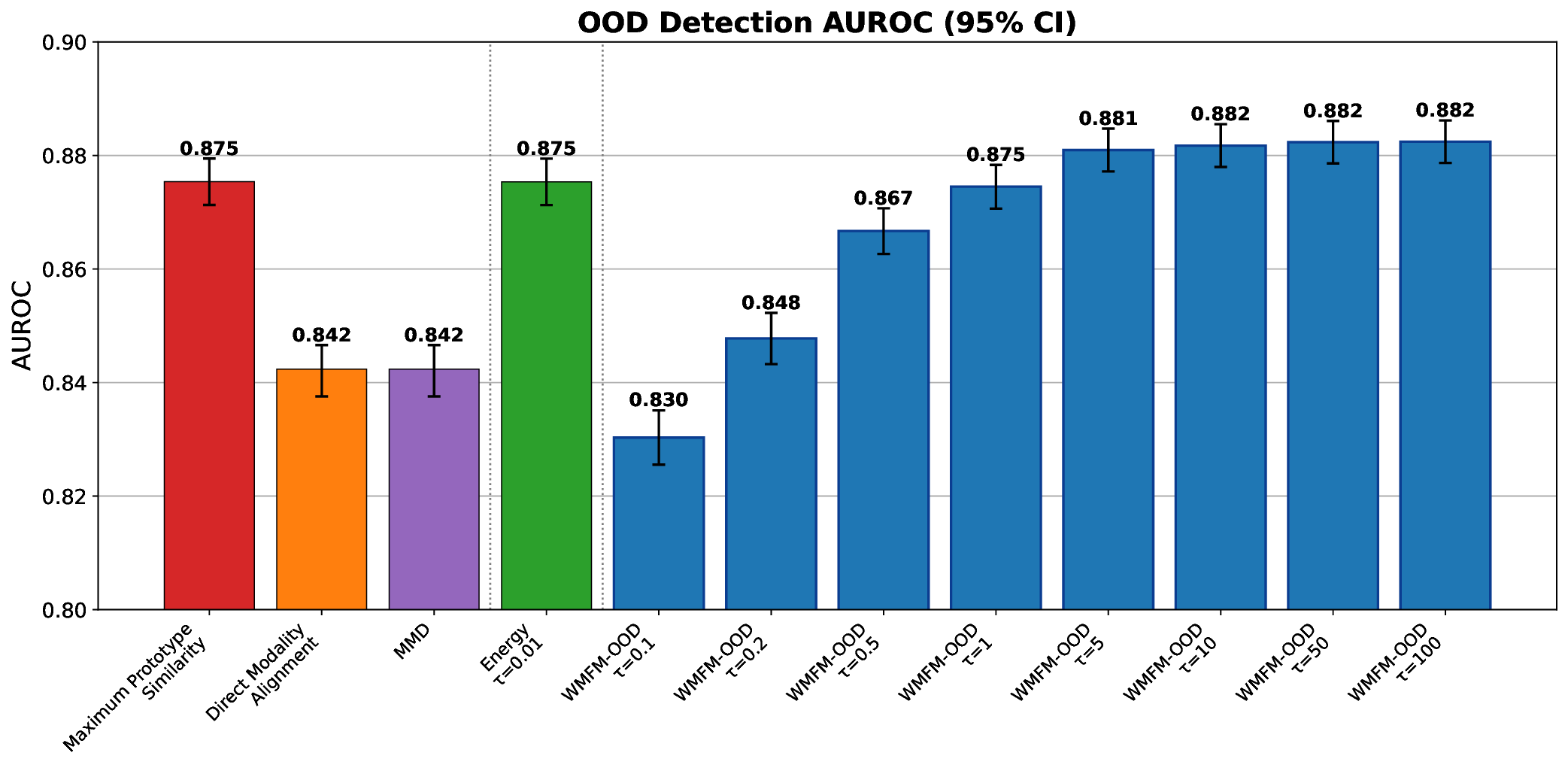}%
        \label{fig:auroc_temp}%
    }%
    \hfil 
    \subfloat[FPR@95\%TPR]{%
        \includegraphics[width=1\linewidth, clip]{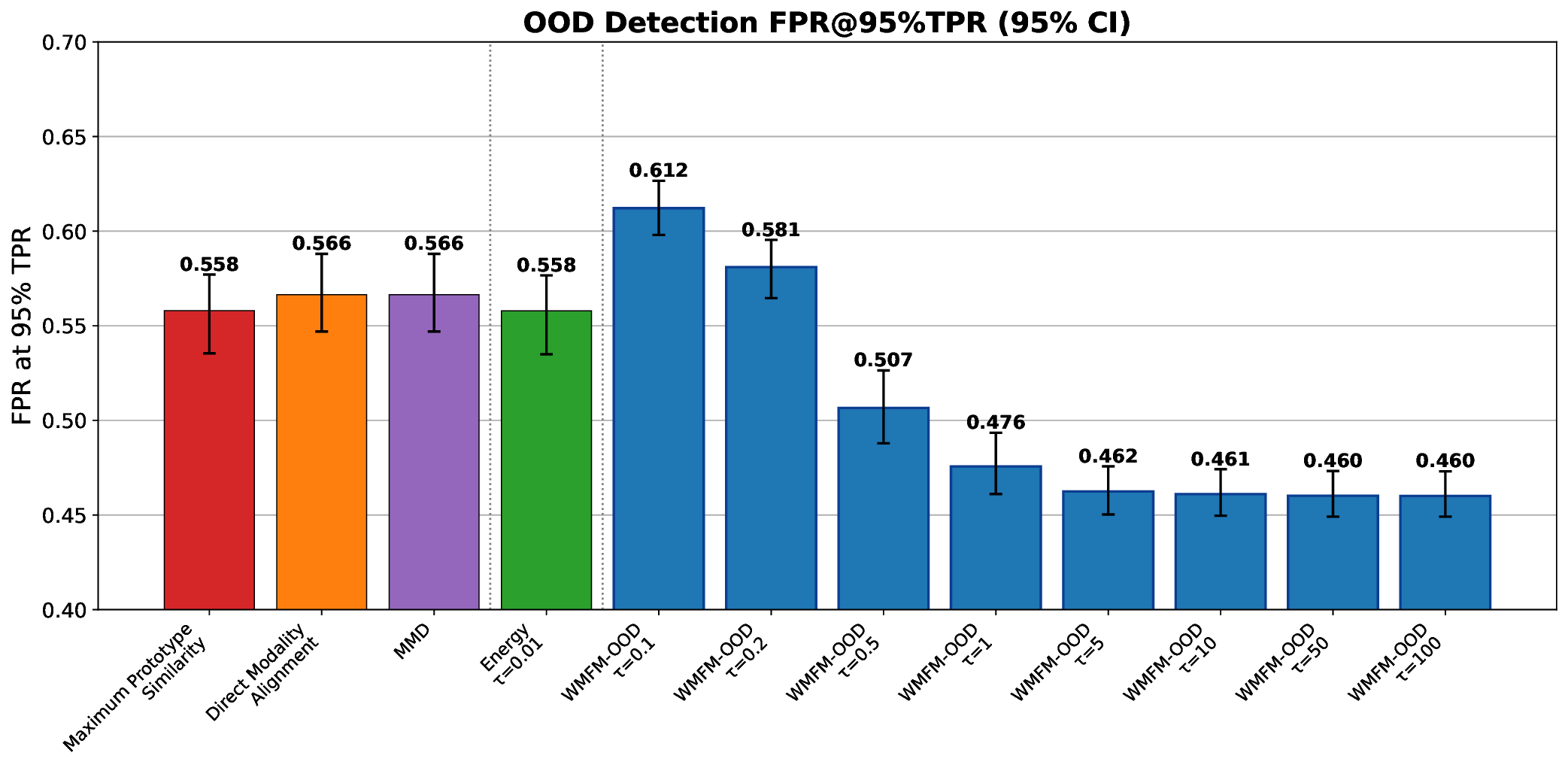}%
        \label{fig:fpr_temp}%
    }
    
    \caption{AUROC and FPR vs Temperature $\tau$.}
    \label{fig:temp_bm_comparison}
\end{figure}

Fig.~\ref{fig:auroc_temp} illustrates the impact of temperature scaling on OOD detection efficacy. While the standalone alignment and discrepancy baselines ($S_{\text{DMA}}$ and $S_{\text{MMD}}$) stall at an AUROC of $0.842$, and the localized baselines ($S_{\text{MPS}}$ and $S_{\text{Energy}}$) saturate at $0.875$, \textbf{WMFM-OOD} achieves superior macro-separability, plateauing at an AUROC of $0.882$ for $\tau \geq 10$. The trend reveals that hard calibration ($\tau \le 0.5$) drastically degrades detection capability, dropping to an operational low of $0.830$ at $\tau = 0.1$ by suppressing useful secondary prototype affinities. In contrast, increasing the temperature produces a robust, soft probability distribution that maximizes the separability between ID and OOD samples. The stable performance plateau and tight confidence intervals at $\tau \geq 10$ further indicate that our method is highly robust to hyperparameter tuning in this linearizing regime.

Fig.~\ref{fig:fpr_temp} depicts the corresponding FPR95 performance (lower is better). While the global alignment and discrepancy baselines stagnate at an elevated false alarm rate of $0.566$, and the non-parametric prototype and energy benchmarks sit at $0.558$, \textbf{WMFM-OOD} significantly suppresses false positives. It compresses the FPR95 down to $0.461$ at $\tau = 10$, reaching an optimal low of $0.460$ at higher temperatures ($\tau \geq 50$). Conversely, low calibration settings ($\tau < 1$) severely exacerbate false alarms, peaking at an nonviable rate of $0.612$ at $\tau = 0.1$ due to the artificial overconfidence scaling of out-of-distribution inputs. Operating in the soft high-temperature regime preserves the localized distributional uncertainty required to reliably isolate environmental shifts, yielding an approximate $17\%$ relative reduction in critical false alarms compared to the uncalibrated prototype baseline.

\begin{figure}
    \centering
    \includegraphics[width=1\linewidth]{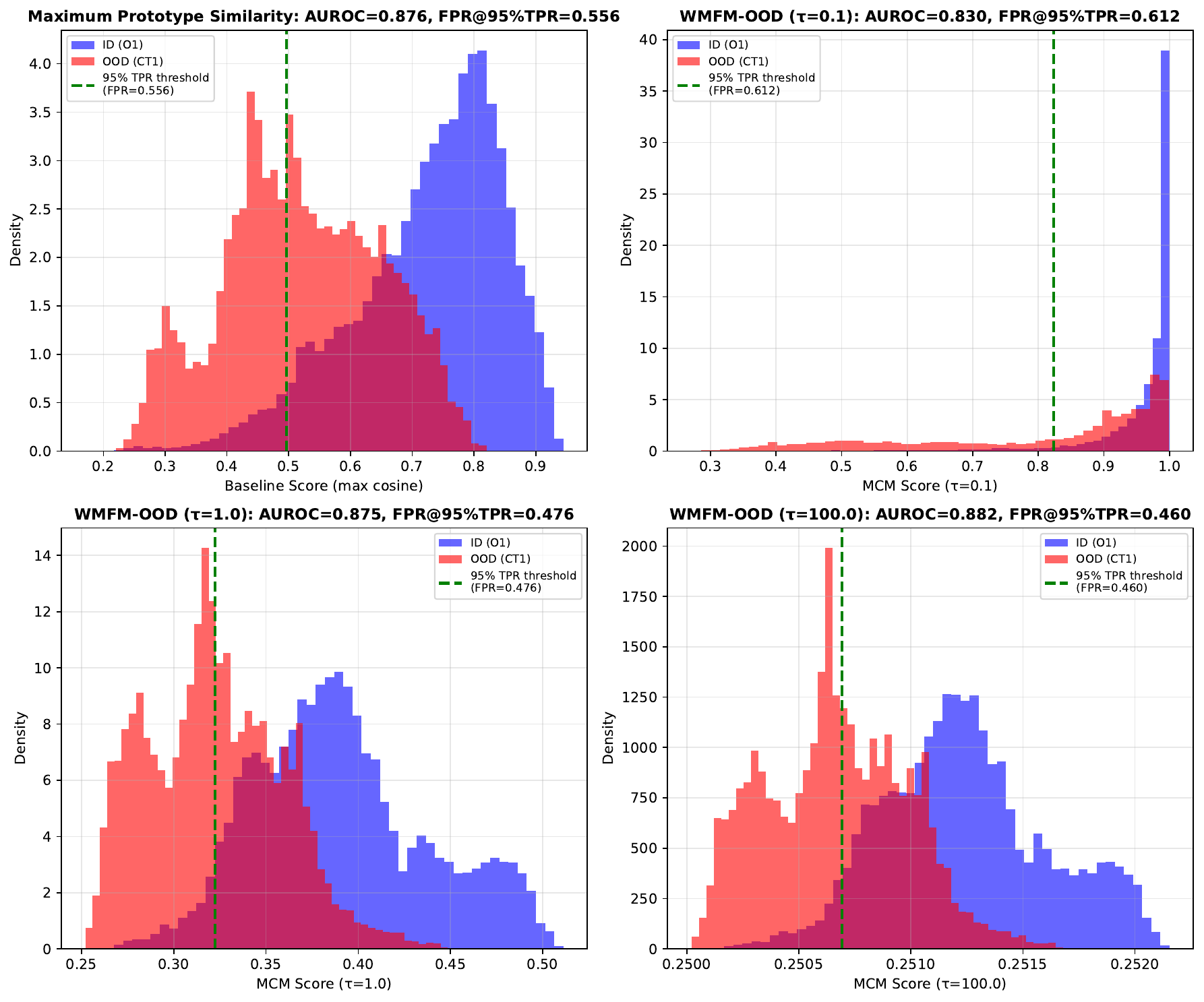}
    \caption{Density histograms of ID and OOD scores.}
    \label{fig:histograms}
\end{figure}

To elucidate the mechanism behind the performance gains, Fig.~\ref{fig:histograms} visualizes the density histograms of detection scores for ID (Blue) and OOD (Red) samples. The analysis reveals that both the uncalibrated baseline and the hard calibration regime ($\tau=0.1$) suffer from significant distributional overlap; specifically, at $\tau=0.1$, the saturating nature of the Softmax function forces OOD samples to mimic high-confidence predictions, rendering them indistinguishable from ID data. In contrast, increasing the temperature to $\tau=100$ effectively softens the probability distribution, revealing the inherent uncertainty in OOD samples. This results in a distinct leftward shift of the OOD density curve away from the ID cluster, thereby creating a cleaner decision margin and minimizing the FPR95 to 0.460.

\subsection{Ablation Study}

\begin{figure}
    \centering
    \subfloat[AUROC]{%
        \includegraphics[width=0.48\linewidth, clip]{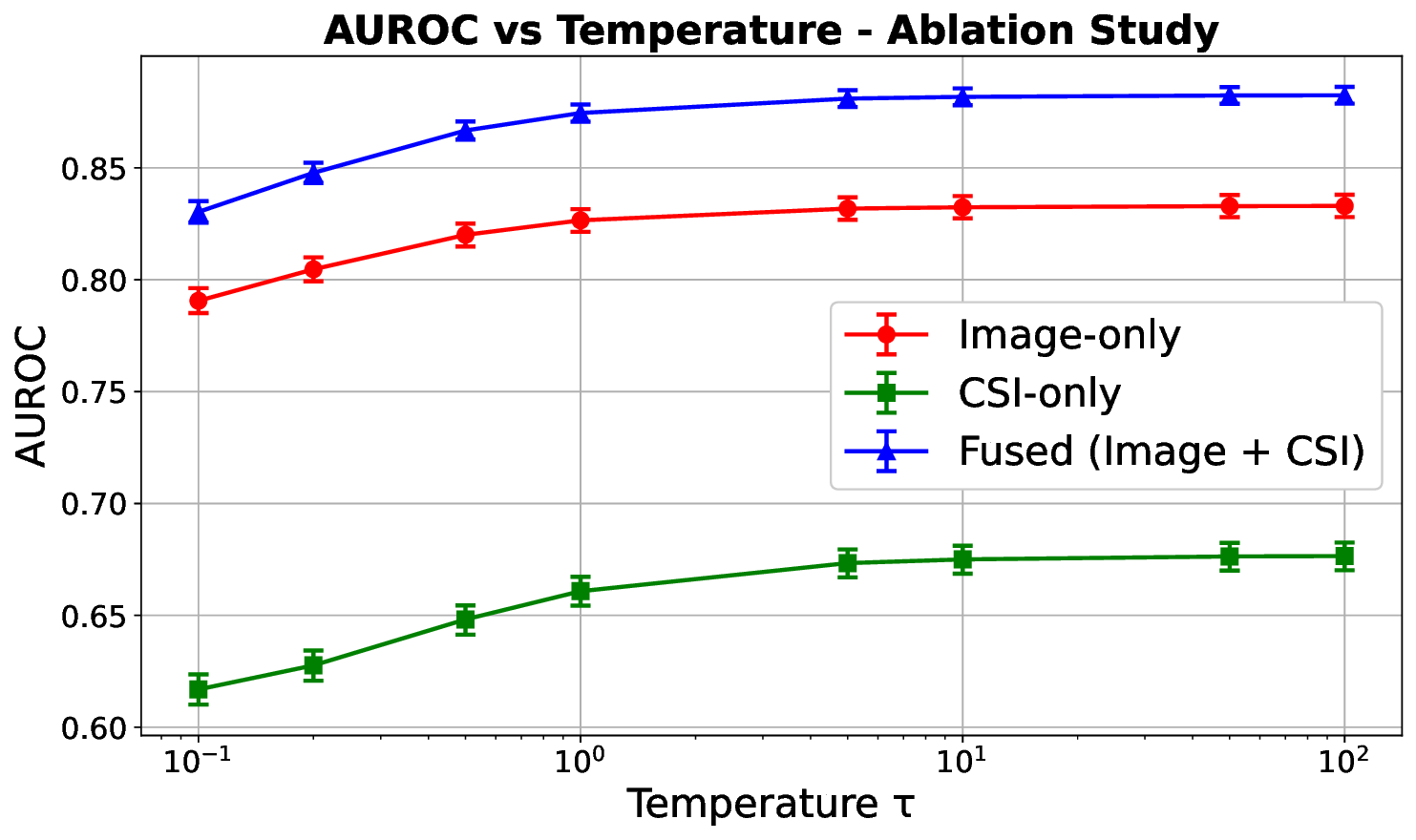}%
        \label{fig:auroc_ablation}%
    }%
    \hfill 
    \subfloat[FPR@95\%TPR]{%
        \includegraphics[width=0.48\linewidth, clip]{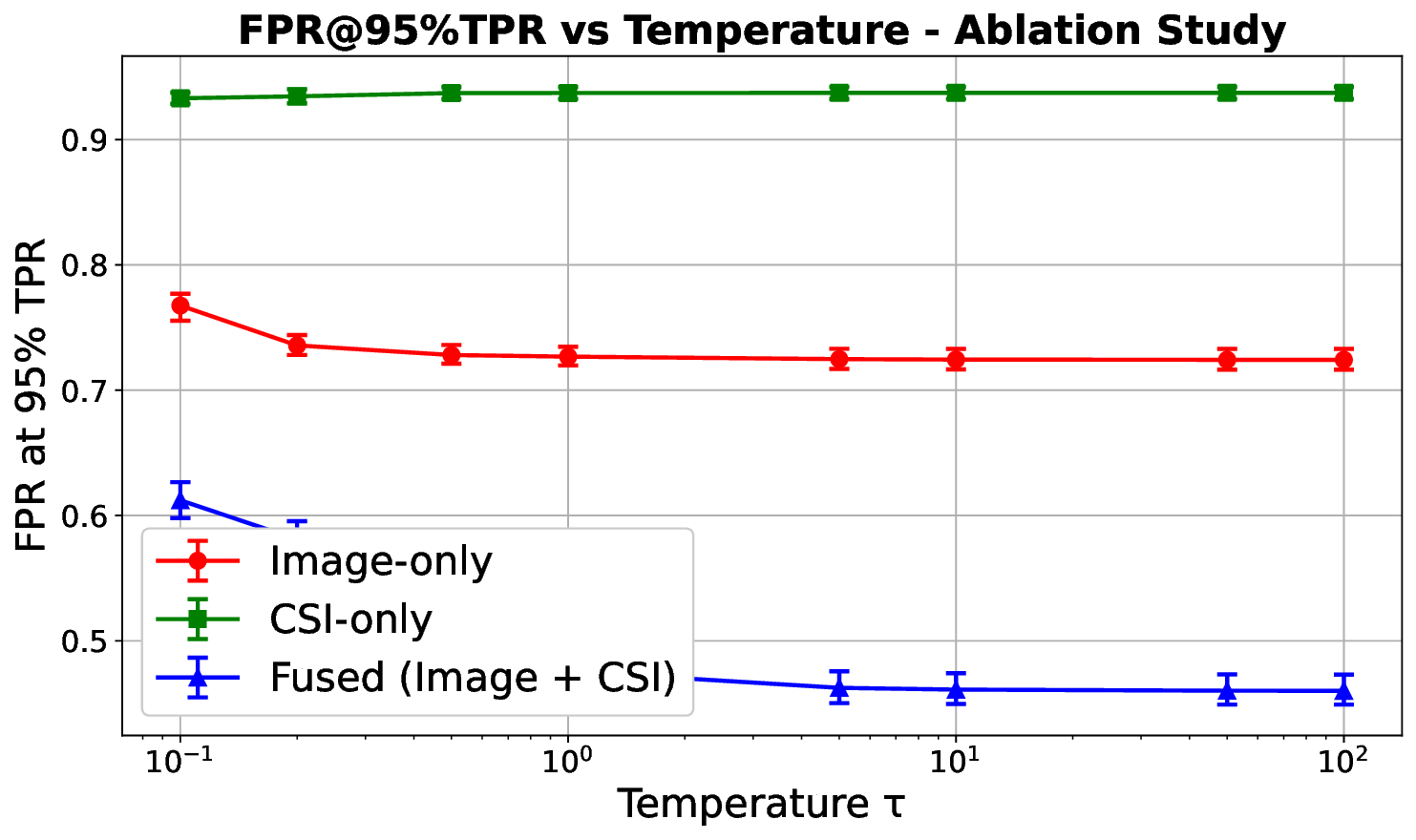}%
        \label{fig:fpr_ablation}%
    }
    
    \caption{Ablation study showing AUROC and FPR95 vs Temperature.}
    \label{fig:ablation}
\end{figure}

To validate the necessity of multimodal integration, Fig.~\ref{fig:ablation} decomposes detection performance across the constituent modalities. The CSI-only baseline (Green) exhibits the lowest discriminative power (AUROC $\approx 0.67$), suggesting that wireless channel features alone are insufficient to fully capture environmental semantics. While the \textbf{Image-only} baseline (Red) provides stronger separation (AUROC $\approx 0.83$), it still suffers from a high false alarm rate (FPR95 $> 0.70$). 
Crucially, the Fused (Image + CSI) framework (Blue) consistently outperforms both single-modality baselines across the entire temperature spectrum. By achieving an AUROC of $0.8824$ and reducing the FPR95 to $\approx 0.46$, the fused approach demonstrates that the synergy between visual context and wireless propagation profiles is essential for robust OOD detection.

\section{Conclusion}
In this paper, we presented \textbf{WMFM-OOD}, a framework safeguarding wireless foundation models against distribution shifts via geometric BS Prototypes and temperature-scaled scoring. Evaluation on the DeepVerse6G dataset demonstrates that explicitly modeling class centroids significantly outperforms raw alignment, while soft calibration ($\tau \geq 10$) reduces the FPR95 to 0.46, a 17\% improvement over baselines. Crucially, this reduction in false alarms is vital for practical deployment, ensuring that the system maintains high safety standards without causing frequent service disruptions. Furthermore, ablation studies confirm that fusing visual and wireless modalities is indispensable for robustness. 
These results establish WMFM-OOD as a promising baseline milestone toward achieving reliable open-set multi-modal architectures in future 6G deployments.

\bibliographystyle{IEEEtran} 
\bibliography{vtc}

\end{document}